\documentclass[preprint,12pt]{elsarticle}

\usepackage{amssymb}

\usepackage{multicol}
\usepackage{subcaption}
\usepackage{float}
\usepackage{siunitx}
\usepackage{lipsum}
\makeatletter
\def\ps@pprintTitle{%
 \let\@oddhead\@empty
 \let\@evenhead\@empty
 \def\@oddfoot{}%
 \let\@evenfoot\@oddfoot}
\makeatother
\usepackage{graphicx}
\usepackage[export]{adjustbox}

\journal{Nuclear Physics B}

\begin{document}

\begin{frontmatter}



\title{Deforestation Prediction Using Neural Networks and Satellite Imagery in a Spatial Information System}


\author{Vahid Ahmadi}

\address{Tarbiat Modares University}
\ead{vahid.ahmadi@modares.ac.ir}
\begin{abstract}
Deforestation, as one of the challenging environmental problems in the world, has been recorded the most serious threat to environmental diversity and one of the main components of land-use change. In this paper, we investigate spatial distribution of deforestation using artificial neural networks and satellite imagery. Modeling deforestation can be conducted considering various factors in determining the relationship between deforestation and environmental and socioeconomic factors. Therefore, in order to ascertain this relationship, the proximity to roads and habitats, fragmentation of the forest, height from sea level, slope, and soil type. In this research, we modeled land cover changes (forests) to predict deforestation using an artificial neural network due to its significant potential for the development of nonlinear complex models. The procedure involves image registration and error correction, image classification, preparing deforestation maps, determining layers, and designing a multi-layer neural network to predict deforestation. The satellite images for this study are of a region in Hong Kong which are captured from 2012 to 2016. The results of the study demonstrate that neural networks approach for predicting deforestation can be utilized and its outcomes show the areas that destroyed during the research period.

\end{abstract}


\end{frontmatter}


\section{Introduction}

Deforestation is an important issue that has received considerable attentions in many different disciplines. As shown in previous studies {cite}, this phenomenon has a negative impact on regional hydrology, large-scale and long-term climatic systems, global biogeochemical cycles, and extinction of animal species \citep{steffen2015planetary}, \citep{allen2015underestimation}, \citep{turner1995land} \citep{howarth2017enhancing} \citep{crowther2014predicting}. Despite its importance, most countries, including Iran, do not have detailed statistics on the extent of deforestation. In Mexico, authors have reported deforestation rates of 0.3-0.5\% per year \citep{richards2014spatially},\citep{newbold2015global},\citep{kibret2016assessment},\citep{faggin2016sustainable},\citep{margono2014primary},\citep{pelletier2017effect}. Also, one study estimated the rate of deforestation to be about 0.3\% and 0.8\% per year for the temperate and tropical areas. This study showed that 84,000 km2 of forest cover were destroyed between 1976 and 2000 \citep{kibret2016assessment}, \citep{pelletier2017effect}. 

Geospatial Information System(GIS) and Remote Sensing are two essential technologies which can be used to implement and couple spatial analyses and machine learning algorithms for predicting deforestation. Spatial analyses cover a wide spectrum of different spatial problems in various scales. It can be used for very small scale movement pattern analysis,  indoor navigation and obstacle detection \citep{nissimov2015obstacle}, \citep{gharani2017context},hydrology, way-finding and path planning \citep{costante2016perception}, large-scale accessibility measurement \citep{zenk2005neighborhood}, \citep{luo2009enhanced},\citep{Gharani2015}, urban informatic, public health, environmental preservation, agriculture, military, and so on. In this research we utilize GIS by integrating it RS and implementing neural network in it. 

Although the main driving factors in deforestation are known, it is difficult to assess their contribution to deforestation and to our best of knowledge, there is no clear understanding of the interaction of these factors. Simulation of land cover and land use change plays a crucial role in management natural resources as well as in academic research. In deforestation, the development of models is accomplished by several useful factors \cite{singh2017modeling}, \cite{liu2014spatiotemporal}, \cite{wood2017consequences}:
Providing a better understanding of how deforestation factors play a role Production of a future scenario for deforestation Forecast of forest destruction In order to support the design of responsible forestry policy The purpose of this study is to develop a simple spatial model that can predict deforestation using artificial neural networks \cite{muller2016long},\cite{pfeifer2017creation},\cite{hansen2013high}.

\section{Combining the potential of deforestation in a database of spatial information systems}

Various factors in determining the relationship between deforestation and environment and socioeconomic factors can be considered. Effective factors such as distance from roads, distance from habitats, altitude, slope, soil gender and patterns of forest fragmentation can be mentioned. Several important factors that describe deforestation include:
\begin{enumerate}

\item Height
\item Slope
\item Soil type
\item The shortest distance from the nearest one
\item The shortest distance from the nearest habitat
\item The shortest distance from the nearest edge of the forest or non-forest
\item The fragmentation of the forest covering the surrounding and immediate area of the site
\end{enumerate}

Forest fragmentation can be estimated using two indicators
\begin{itemize}

\item Forest cover index
\item Matron index

\end{itemize}

The forest cover index, expressed as percentages, shows the calculated pixel count of the forest in $3 \times 3$, $9 \times 9$ or $15 \times 15$ pixels windows. In this study, a $3 \times 3$ pixel window was used. The matron index is defined in the $3 \times 3$ pixel window as follows:
\begin{equation}
M=\frac{N_{F-NF}}{\sqrt{N_F}\sqrt{N}}
\end{equation}
Where $NF$ is the number of boundary pixels between forest and non-forest, NF is the number of pixels in the forest and N is the total number of pixels \citep{mertens1997spatial}.

To determine the number of suitable variables for entering the artificial neural network, the correlation coefficients between the variables can be calculated \citep{mather2016classification},\citep{biehl2002multispec}. Variables such as altitude, forest cover and distance and proximity to residential areas were selected as network inputs.
\section{Multilayer Perceptron Neural Network}

Selected variables were used to describe the network's training data. In the model, the data were divided into two categories: training set and test set. The training algorithm does not use under any circumstances the test data to adjust network weights because the test set is used to detect network execution errors and stop network training if over-learning occurs, so there should be no dependence on network parameters.

We used a neural network with similar configuration to \citep{gharani2017artificial} in which the training process was tested with Levenberg-Marquadt and Back-propagation algorithms. Deciding on the number of hidden layers experimentally by testing various choices that provide the best balance between the bias and the variance \citep{geman1992neural}, \citep{lake2017building}, \citep{bishop1995neural}. Finally, a network with three levels of input layer, hidden layer and output layer was designed (\textbf{Figures 1}).

\begin{figure}[H]
\begin{center}

\includegraphics[scale=.7]{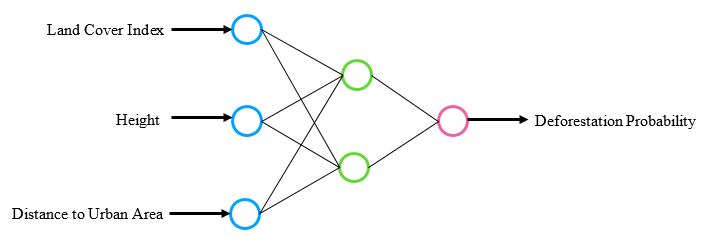}
\caption{Schematic diagram of the Network}

\end{center}
\label{figure1}
\end{figure}

Using layers in GIS environment, the network was trained with existing data and then used to prepare a deforestation risk map. The network output is a quantity that expresses the natural tendency of each pixel for deforestation. So, the result is a fuzzy deforestation map that shows the probability of deforestation, which ultimately is classified.

\section{The study area and the stages of the work}
The studied area is part of the forested area in Hong Kong, China, which is located in the southeast of Asia between $N\ang{22;4;}$--$N\ang{22;43;}$  and $E\ang{113;4;}$--$E\ang{114;36;}$ ( \textbf{Figures 2}).

\begin{figure}[H]
\begin{center}

\includegraphics[scale=.6]{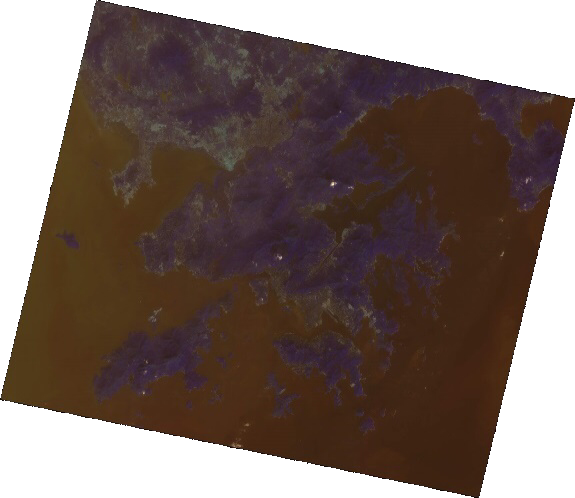}
\caption{Satellite image of the study area}

\end{center}
\label{figure2}
\end{figure}
Two SPOT images of the same area, taken on 14/01/2012 and 20/12/2016, were used along with ground control points for geometric correction and regression, and a 90-meter DEM of the region, which was provided by the SRTM database. Stages of work in the model The study consists of six steps as follows:
\begin{enumerate}

\item Registry and image correction
\item Classification of images into three classes of forest, wetland and urban areas
\item Preparation of deforestation maps obtained from overlapping forest cover maps multiple times
\item Obtaining appropriate layers and their configuration for modeling
\item Modeling (Neural Network Training)
\item Simulation (providing a natural tendency map for deforestation that predicts deforestation for the next period)

\end{enumerate}

To carry out the steps, first, the images taken in the ArcGIS environment were retrieved using Extention, Georefrencing, as shown in \textbf{Figures 3 (a) and 3 (b)}.

\begin{figure}[H]
\begin{multicols}{2}
    \includegraphics[width=\linewidth,frame]{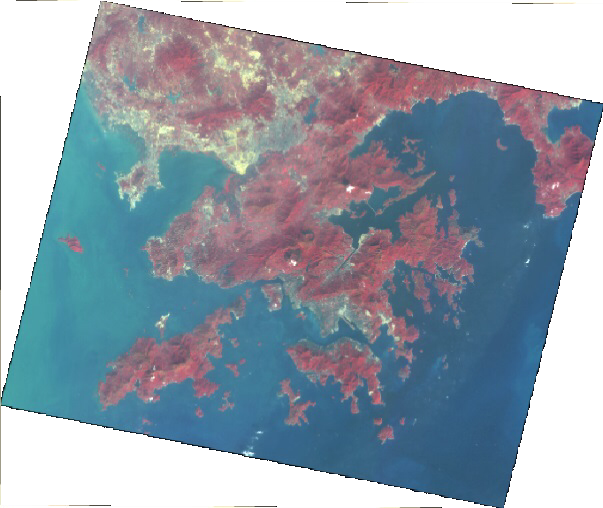}\par 
    \subcaption{2012}
    \includegraphics[width=\linewidth,frame]{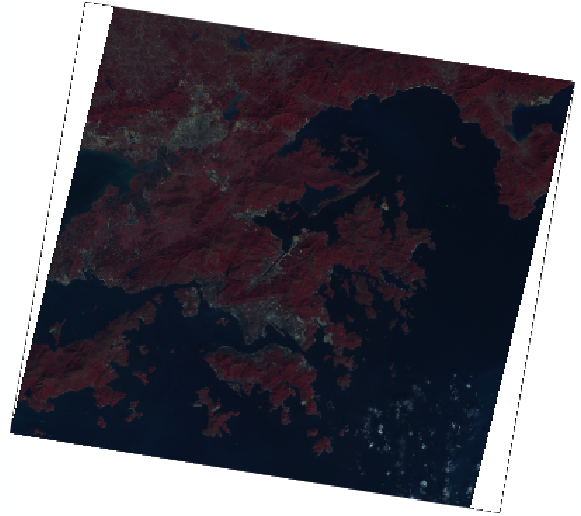}\par 
    \subcaption{2016}
    
    \end{multicols}
    \caption{Georeferenced SPOT satellite images of the area in two snapshots}
\end{figure}
The images were then embedded in the ERDAS software and the images were classified using the algorithm most similarly to forest, sea, and urban classes and loaded again into the ArcGIS environment. The clustered images are shown in \textbf{Figures 4(a) and 4(b)}

\begin{figure}[H]
\begin{multicols}{2}
    \includegraphics[width=\linewidth,frame]{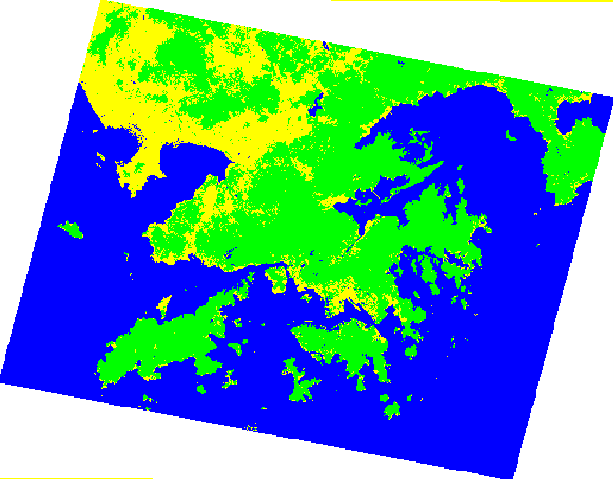}\par 
    \subcaption{2012}
    \includegraphics[width=\linewidth,frame]{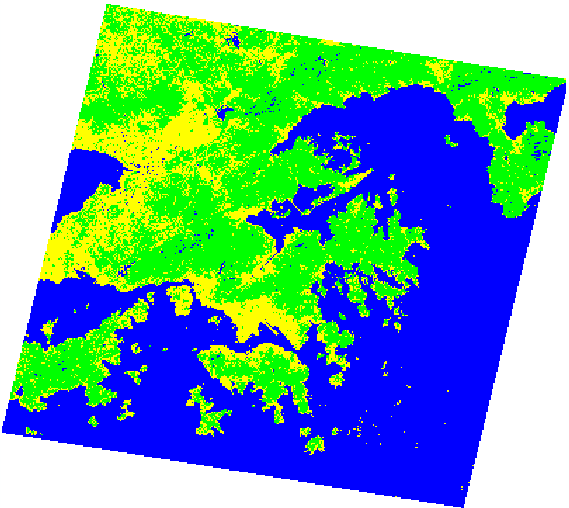}\par 
    \subcaption{2016}
    
    \end{multicols}
    \caption{Classified images into forest, sea, and urban areas}
\end{figure}
From Fig. 4 (a) and Fig. 4 (b), a layer is derived for forest areas. In the output images, only two values are zero (forests) and one (non-forest). The output of this process for the two years studied is shown in Figures 5 (a) and 5 (b):

\begin{figure}
\begin{multicols}{2}
    \includegraphics[width=\linewidth,frame]{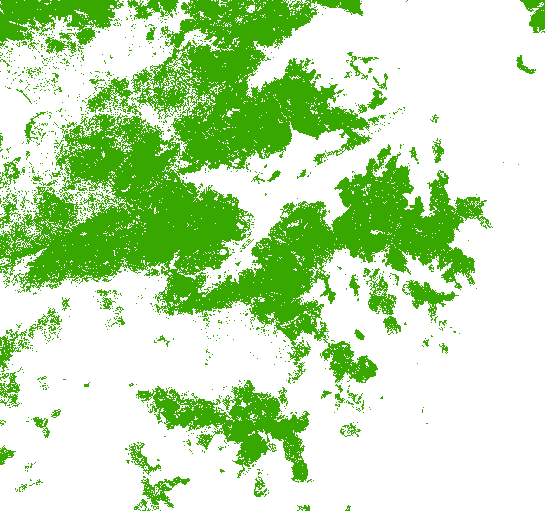}\par 
    \subcaption{2012}
    \includegraphics[width=\linewidth,frame]{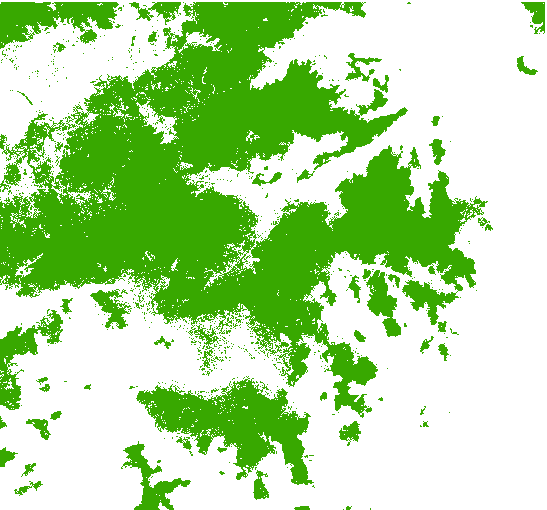}\par 
    \subcaption{2016}
    
    \end{multicols}
    \caption{Reclassification into forest and non-forest classes}
\end{figure}
These layers were used to generate the forest cover index. For this purpose, these layers were outputted in ASCII format and MATLAB software was written for this purpose, and its output was returned to the ArcGIS environment again. This output resulted in two images depicted in Figures 6 (a) and 6 (b).

\begin{figure}
\begin{multicols}{2}
    \includegraphics[width=\linewidth,frame]{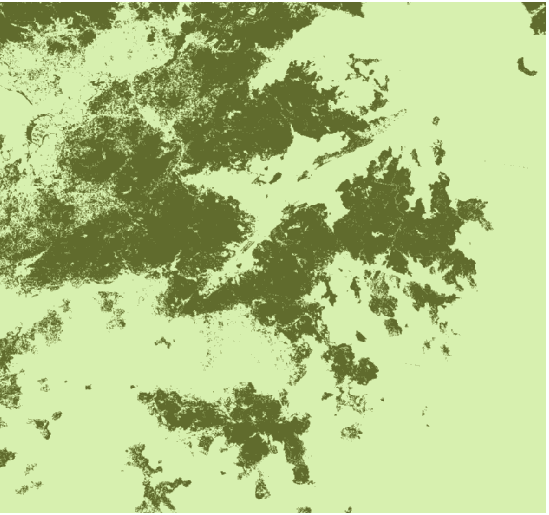}\par 
    \subcaption{2012}
    \includegraphics[width=\linewidth,frame]{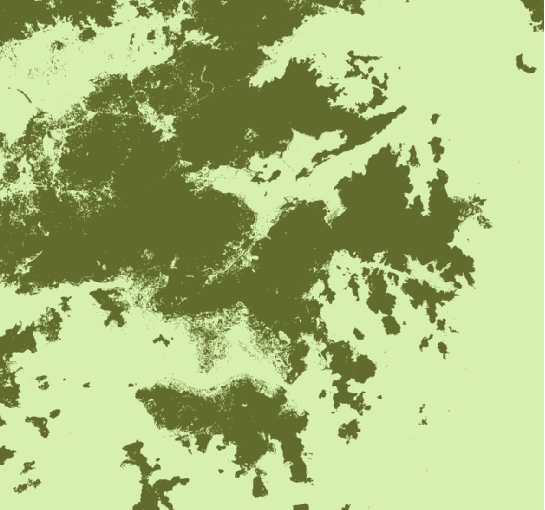}\par 
    \subcaption{2016}
    
    \end{multicols}
    \caption{Generated forest cover index for 2012 and 2016}
\end{figure}
The two layers of the forest cover index are the first line of entry for the network. Now the production of the second layer, distance and proximity to the cities, has gone as the second line of input of the network. For this purpose, a layer is first produced which consists of two classes of urban and non-urban, the image of these two classes in Fig. 7 (a) and Fig. 7 (b).

\begin{figure}
\begin{multicols}{2}
    \includegraphics[width=\linewidth,frame]{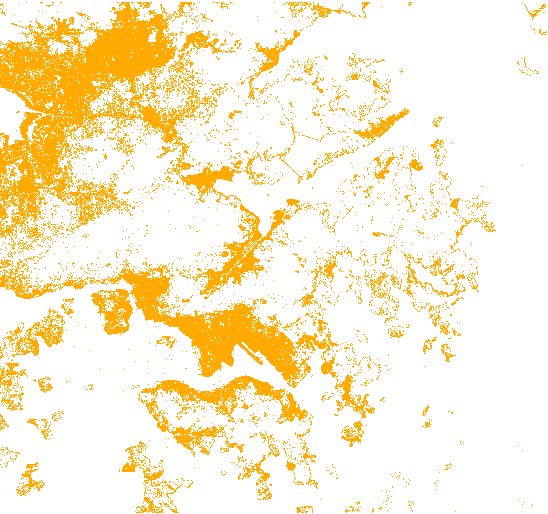}\par 
    \subcaption{2012}
    \includegraphics[width=\linewidth,frame]{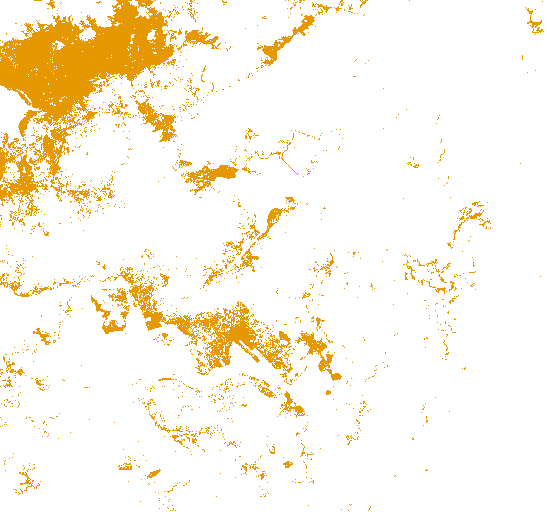}\par 
    \subcaption{2016}
    
    \end{multicols}
    \caption{Reclassification into urban and non-urban classes}
\end{figure}
Now with the help of the Spatial Analyst functions in ArcGIS software, we produce the layer for entering the network, distance from urban areas. In Figure 8 (a) and Figure 8 (b), the image of this layer is well visible.

\begin{figure}
\begin{multicols}{2}
    \includegraphics[width=\linewidth,frame]{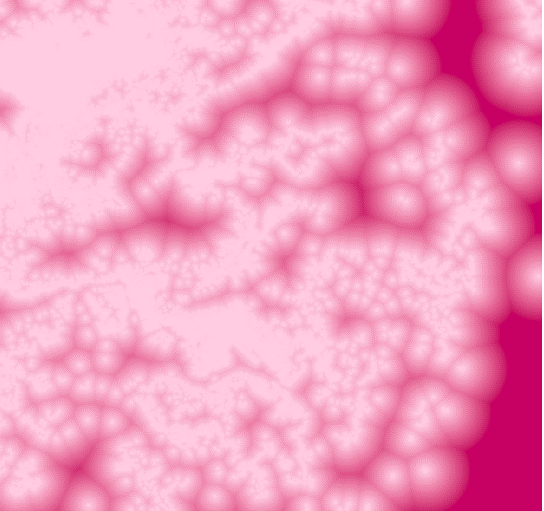}\par 
    \subcaption{2012}
    \includegraphics[width=\linewidth,frame]{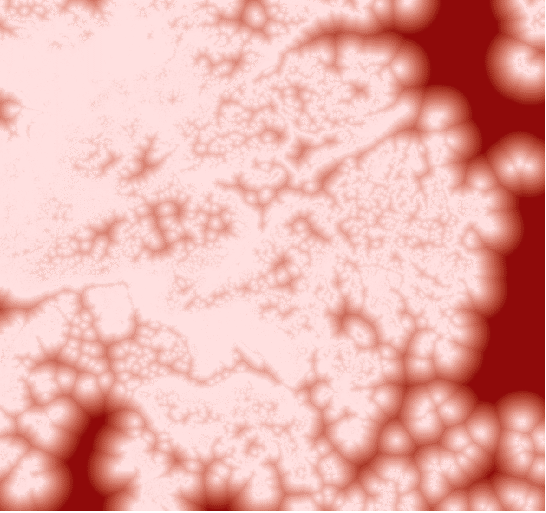}\par 
    \subcaption{2016}
    
    \end{multicols}
    \caption{Distance layer - computed distance from urban areas }
\end{figure}
Finally, the last entry as the third line of the vector input is the height. As mentioned earlier, the region DEM (Fig. 9) was used after a series of processes to extract a common area. This DEM contains information in non-marine areas with elevations ranging from 943 to 80 m.

\begin{figure}[H]
\begin{center}

\includegraphics[scale=.5,frame]{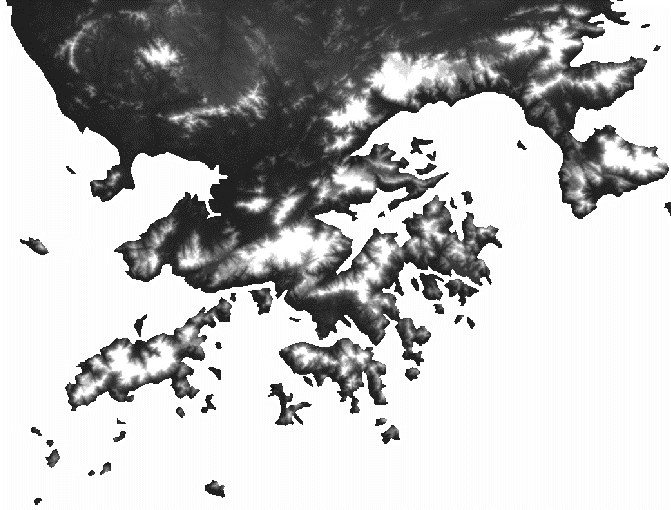}
\caption{Schematic diagram of the Network}

\end{center}
\label{figure1}
\end{figure}

\begin{figure}[H]
\begin{center}

\includegraphics[scale=.5,frame]{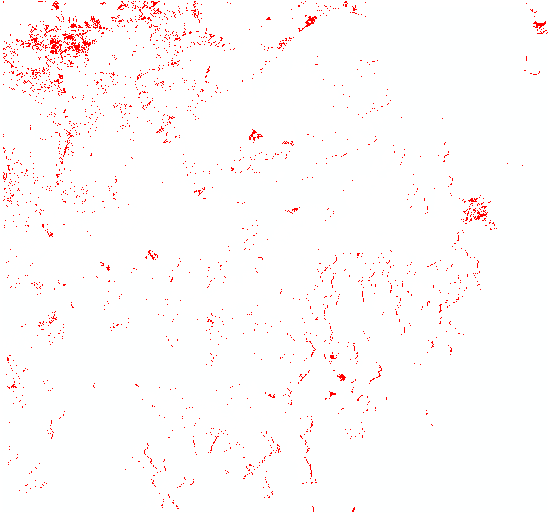}
\caption{Schematic diagram of the Network}

\end{center}
\label{figure1}
\end{figure}
Once the network input data is ready, the training data must be generated. For this purpose, there has to be a process to explore deforestation during these years. In this process, the areas covered by forest cover in 2012 and in 2016 will be covered outside of the forest, amounting to one, and if they maintain their forest cover, zero is allocated to non-refugee areas in 2012, NoData will be allocated. The output of this process is shown in Figure -10.
By taking the ASCII output from this map, which indicates actual deforestation, part of it, along with the corresponding entries for 2012, is given as training data for network education. The other part of the data that was not involved in the training process was used to test the network, resulting in a precision of 98
Network convergence was performed by setting the network parameters as follows to achieve the desired accuracy. (Figure-11)

\begin{figure}[H]
\begin{center}

\includegraphics[scale=.5,frame]{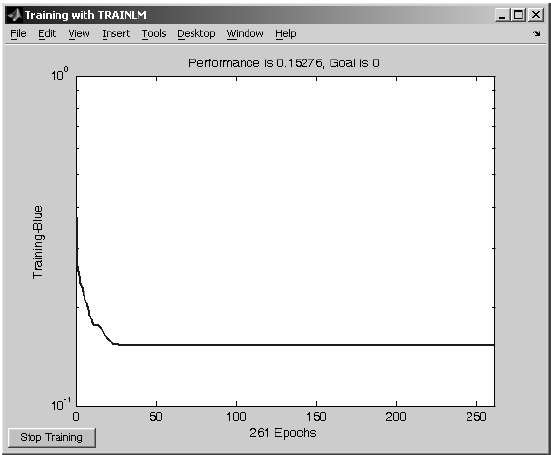}
\caption{Schematic diagram of the Network}

\end{center}
\label{figure1}
\end{figure}

\begin{figure}[H]
\begin{center}

\includegraphics[scale=.5,frame]{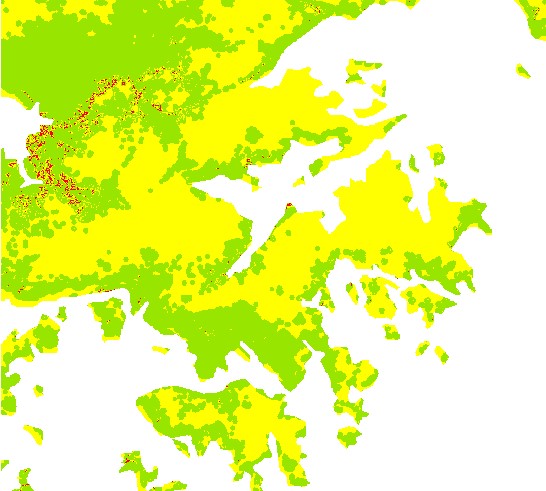}
\caption{Schematic diagram of the Network}

\end{center}
\label{figure1}
\end{figure}
Once the network parameters are set, it is simulated. To do this, we have to retrieve the data from 2016 to the network to get results about the future.
The result of this stage is a de industrial risk map divided into three classes, class one related to areas without deforestation, the second class relates to areas relatively prone to deforestation and to some extent sustainable, and the third class relates to areas at high risk of deforestation and Extremely unstable.
In Figure 12, a final map of the prediction of deforestation in the studied area is presented.





\section{Conclusion}

In this paper, the ability of artificial neural networks to model land cover changes as a powerful method for investigating deforestation was investigated. However, the development of models for deforestation and land use change due to the high dependence of models on large environmental factors, due to changing climatic and economic conditions, is practically impossible to accurately predict.
Therefore, in this research, using the neural networks of a part of the forest that was destroyed was accurately predicted. This model can be used by environmentalists and managers to develop targeted policies to control the ecological and social impacts of deforestation.

  \bibliographystyle{elsarticle-num}

\end{document}